\documentclass[12pt]{article}
\usepackage{bm}
\begin{document}
\title{Truly Minimal Left-Right Symmetry Model for Electroweak Interaction}
\date{}
\maketitle
\begin{center}
\textbf{Asan Damanik\footnote{E-mail: d.asan@lycos.com}}\\
Department of Physics, Sanata Dharma University,\\ Kampus III USD Paingan, Maguwoharjo, Sleman,Yogyakarta, Indonesia\\
and Graduate School, Gadjah Mada University, Yogyakarta,Indonesia\\
\end{center}
\begin{center}
\textbf{Mirza Satriawan} and \textbf{Arief Hermanto}\\
Department of Physics, Gadjah Mada University, Yogyakarta, Indonesia\\
\end{center}
\begin{center}
\textbf{Pramudita Anggraita}\\
National Nuclear Energy Agency (BATAN), Jakarta, Indonesia
\end{center}     
\abstract{By using two primary doublets and one induced bidoublet Higgs fields as a result of the interactions of the two doublets, we evaluate the predictive power of the left-right symmetry model based on $SU(2)_{L}\otimes SU(2)_{R}\otimes U(1)$ gauge group to the gauge bosons masses, leptons masses, and the structure of electroweak interactions. We found that the contribution of the right charge-current to the electroweak interaction is only around 0.0073 percent. The neutrino mass emerges naturally without introducing exotic particles.  We obtain that the mixing angle $\theta = 45^{0}$ for boson sector, $\theta \approx 45^{0}$ for neutrino sector, and $\theta = 0^{0}$ for electron sector.  The parity violation in our model could be associated with the mass mixing in the bosons and leptons masses arise from the induced bidoublet Higgs via symmetry breaking.}

\begin{flushleft}
PACs: 12.60.Cn; 12.60.Fr
\end{flushleft}

\section{Introduction}
Even though the Glashow-Weinberg-Salam (GWS) model for electroweak interaction based on $SU(2)_{L}\otimes U(1)_{Y}$ gauge group very successful phenomenologically \cite{Collins89, Gross93, Fukugita03}, but many fundamental problems i.e. the responsible mechanism for generating the neutrino masses, the domination of the V-A over V+A interactions at low energy, and the responsible mechanism for the doublet lepton up-down mass difference could not be explained by GWS model.  The need for the extension of the GWS model comes from the conclusion that the neutrinos have a mass as a direct implication of the detected neutrino oscillations phenomena for both solar and atmospheric neutrinos \cite{Fukuda98, Fukuda99, Fukuda01, Toshito01, Giacomelli01, Ahmad02, Ahn03}.
Many theories or models to extend the GWS model have been proposed.  One of the interesting models is the Left-Right symmetry model based on $SU(2)_{L}\otimes SU(2)_{R}\otimes U(1)$ gauge group proposed by  many authors \cite{Pati75, Mohapatra75a, Senjanovic75, Mohapatra75, Senjanovic79, Mohapatra81, Simoes02}.  In this paper, we use the left-right symmetry model based on $SU(2)_{L}\otimes SU(2)_{R}\otimes U(1)$ gauge group with two doublets and one bidoublet Higgs fields (as a result of the two doublets interaction) to break the $SU(2)_{L}\otimes SU(2)_{R}\times U(1)$ gauge group down to $U(1)em$.  The lepton fields to be represented as doublet of an SU(2) for both left and right fields to make the full sense of the left-right symmetry model.
The paper is organized as follows: in section 2 we present our main assumptions for the left-right symmetry model, in section 3 we evaluate the gauge bosons masses, in section 4 we evaluate the leptons masses, and finally in section 5 we give a conclusion.

\section{The Model}
In our model, the left-right symmetry model for electroweak interaction based on the $SU(2)_{L}\otimes SU(2)_{R}\otimes U(1)$ gauge group with the assumptions and the particles assignment as follows:
\begin{enumerate}
 \item{Two primary Higgs fields are doublet of   SU(2):}
 \begin{eqnarray}
X_{L}=\bordermatrix{&\cr
&a^{+}\cr
&b^{0}\cr},X_{R}=\bordermatrix{&\cr
&c^{+}\cr
&d^{0}\cr}.
 \label{eq:lrhiggs}
\end{eqnarray} 
\item{The secondary bidoublet Higgs field could be produced from the interaction of the two primary Higgs fields, that is:}
\begin{eqnarray}
\phi=\bordermatrix{& &\cr
&p^{0} &q^{+}\cr
&r^{-} &s^{0}\cr}
 \label{eq:phi}
\end{eqnarray}
 \item{The leptons fields are doublet of  SU(2) for both left and right fields:}
\begin{eqnarray}
 \psi_{L}=\bordermatrix{&\cr
&\nu\cr
&e^{-}\cr}_{L},\psi_{R}=\bordermatrix{&\cr
&\nu\cr
&e^{-}\cr}_{R}.
  \label{eq:leptonfield}
\end{eqnarray}
 \item{Both leptons and gauge bosons masses are generated via symmetry breaking when Higgs fields develop its vacuum expectation values similar to the symmetry breaking in the GWS model.}
\end{enumerate}

By applying the above assumptions and particles assignment, we could break the left-right symmetry model based on $SU(2)_{L}\otimes SU(2)_{R}\otimes U(1)$ down to $U(1)em$ directly as shown schematically below:

\begin{center}
$SU(2)_{L}\otimes SU(2)_{R}\otimes U(1)$\\
$\Downarrow$\\
$U(1)_{em}$
\end{center}

The vacuum expectation values of two doublets and one bidoublet Higgs could contribute to the gauge bosons masses.  Meanwhile, the leptons masses (Yukawa term) only come from the vacuum expectation value of bidoublet Higgs, because the leptons fields  to be doublet of $SU(2)$, then only the Yukawa term with bidoublet Higgs satisfies gauge invariance.   Thus, the complete Lagrangian density could be reads:
\begin{eqnarray}
L=-\frac{1}{4}W_{\mu\nu L}.W^{\mu\nu L}-\frac{1}{4}W_{\mu\nu R}.W^{\mu\nu R}-\frac{1}{4}B_{\mu\nu}B^{\mu\nu}\nonumber\\+\bar{\psi_{L}}\gamma^{\mu}\left(i\partial_{\mu}-g\frac{1}{2}\tau.W_{\mu L}-g'\frac{Y}{2}B_{\mu}\right)\psi_{L}\nonumber\\+\bar{\psi_{R}}\gamma^{\mu}\left(i\partial_{\mu}-g\frac{1}{2}\tau.W_{\mu R}-g'\frac{Y}{2}B_{\mu}\right)\psi_{R}\nonumber\\+\left|\left(i\partial_{\mu}-g\frac{1}{2}\tau.W_{\mu L}-g'\frac{Y}{2}B_{\mu}\right)X_{L}\right|^2\nonumber\\+\left|\left(i\partial_{\mu}-g\frac{1}{2}\tau.W_{\mu R}-g'\frac{Y}{2}B_{\mu}\right)X_{R}\right|^2\nonumber\\+Tr\left|\left(i\partial_{\mu}-g\frac{1}{2}\tau.W_{\mu L}-g'\frac{Y}{2}B_{\mu}\right)\phi\right|^2\nonumber\\+Tr\left|\left(i\partial_{\mu}-g\frac{1}{2}\tau.W_{\mu R}-g'\frac{Y}{2}B_{\mu}\right)\phi\right|^2\nonumber\\-V(X_{L},X_{R},\phi)-(G\bar{\psi_{L}}\phi\psi_{R}+H.c.)\ \ \ \ \ \ \ \ 
 \label{eq:completelagrangian}
\end{eqnarray}
where $g_{L}=g_{R}=g$ are the $SU(2)$ couplings, $g'$ is the $U(1)$ coupling, $\gamma^{\mu}$ are the Dirac matrices, $\tau$'s are the Pauli spin matrices, $V(X_{L},X_{R},\phi)$ is the Higgs potential, $Y$ is the hypercharge ($Y=B-L$), and $G$ is the Yukawa coupling.

The electric charge operator $Q$ satisfies the relations:
\begin{eqnarray}
Q=T_{3L}+T_{3R}+\frac{Y}{2}
 \label{eq:chargeop}
\end{eqnarray}
where $T_{3L}$ and $T_{3R}$ are the third components the weak isospin generator: $T_{i}=\frac{\tau_{i}}{2}$.
According to the Eqs. (\ref{eq:completelagrangian}) and (\ref{eq:chargeop}), the Higgs fields have the following quantum numbers:
\begin{eqnarray}
X_{L}\left(\frac{1}{2},0,1\right), X_{R}\left(0,\frac{1}{2},1\right), \phi\left(\frac{1}{2},\frac{1}{2},0 \right)
\end{eqnarray}
and the leptons fields transform as:
\begin{eqnarray}
\psi_{L}\left(2,0,0\right), \psi_{R}\left(0,2,0\right),
\end{eqnarray}
uder $SU(2)$.
\section{The Gauge Bosons Masses}
From Eq. (\ref{eq:completelagrangian}), we can see that the relevant gauge boson mass terms as follow:
\begin{eqnarray}
L_{boson}=\left|\left(-g\frac{1}{2}\tau.W_{\mu L}-g'\frac{Y}{2}B_{\mu}\right)X_{L}\right|^2+\left|\left(-g\frac{1}{2}\tau.W_{\mu R}-g'\frac{Y}{2}B_{\mu}\right)X_{R}\right|^2\nonumber\\+Tr\left|\left(-g\frac{1}{2}\tau.W_{\mu L}-g'\frac{Y}{2}B_{\mu}\right)\phi\right|^2+Tr\left|\left(-g\frac{1}{2}\tau.W_{\mu R}-g'\frac{Y}{2}B_{\mu}\right)\phi\right|^2
 \label{eq:boson1}
\end{eqnarray}
By substituting the vacuum expectation values of the Higgs fields: 
\begin{eqnarray}
\left\langle X_{L}\right\rangle=\bordermatrix{&\cr
&0\cr
&b\cr},\;\left\langle X_{R}\right\rangle=\bordermatrix{&\cr
&0\cr
&d\cr},\; \left\langle \phi\right\rangle=\bordermatrix{& &\cr
&p &0\cr
&0 &s\cr}
 \label{eq:vevhiggs}
\end{eqnarray} 
into Eq.(\ref{eq:boson1}), we obtain:
\begin{eqnarray}
L_{boson}=\frac{g^2b^2}{4}\left\{\left(W_{\mu L}^{1}\right)^2 +\left(W_{\mu L}^{2}\right)^2\right\}+\frac{b^2}{4}\left(gW_{\mu L}^{3}-g'B_{\mu}\right)^2\nonumber\\+\frac{g^2d^2}{4}\left\{\left(W_{\mu R}^{1}\right)^2+\left(W_{\mu R}^{2}\right)^2\right\}+\frac{d^2}{4}\left(gW_{\mu R}^{3}-g'B_{\mu}\right)^2\nonumber\\+\frac{g^2}{4}\left(p^2+s^2\right)\left\{\left(W_{\mu L}^{1}\right)^2+\left(W_{\mu L}^{2}\right)^2+\left(W_{\mu L}^{3}\right)^2\right\}\nonumber\\+\frac{g^2}{4}\left(p^2+s^2\right)\left\{\left(W_{\mu R}^{1}\right)^2+\left(W_{\mu R}^{2}\right)^2+\left(W_{\mu R}^{3}\right)^2\right\}
 \label{eq:lagrangian2}
\end{eqnarray}
By defining:
\begin{eqnarray}
W_{\alpha}^{\pm}=\frac{1}{\sqrt{2}}\left(W_{\mu \alpha}^{1}\mp iW_{\mu \alpha}^{2}\right), Z_{\mu \alpha}=\frac{gW_{\mu \alpha}^{3}-g'B_{\mu \alpha}}{\sqrt{g^2+g'^2}},\\ A_{\mu \alpha}=\frac{g'W_{\mu \alpha}^{3}+gB_{\mu \alpha}}{\sqrt{g^2+g'^2}}, Z_{\mu \alpha}^{'}=W_{\mu \alpha}^{3},\ \ \ \ \ \ \ \ \ \ \ 
 \label{eq:boson}
\end{eqnarray}
where $\alpha=L,R$, and
\begin{eqnarray}
m_{W_{L}}=\frac{gb}{2}, m_{W_{R}}=\frac{gd}{2}, m_{A}=0, m_{Z_{L}}=\frac{b\sqrt{g^2+g'^2}}{2}, m_{Z_{R}}=\frac{d\sqrt{g^2+g'^2}}{2},\nonumber \\ m_{W_{L}}^{*}=m_{W_{R}}^{*}=m_{Z_{L}^{'}}=m_{Z_{R}^{'}}=\frac{g\sqrt{p^2+s^2}}{2},\ \ \ \ \ \ \ \ \ \ \ \ \ \ \ 
 \label{eq:bosonmass}
\end{eqnarray}
then Eq.(\ref{eq:lagrangian2}) reads:
\begin{eqnarray}
L_{boson}=\left(m_{W_{L}}^2+m_{W_{L}}^{*2}\right)W_{L}^{+}W_{L}^{-}+\left(m_{W_{R}}^2+m_{W_{R}}^{*2}\right)W_{R}^{+}W_{R}^{-} \ \ \ \ \ \ \ \ \  \nonumber \\ +m_{Z_{L}}^2Z_{\mu L}Z_{L}^{\mu}+m_{Z_{R}}^2Z_{\mu R}Z_{R}^{\mu}+m_{A}^2A_{\mu}A^{\mu}\nonumber \\+m_{Z_{L}^{'}}^2Z_{\mu L}^{'}Z_{L}^{'\mu}+m_{Z_{R}^{'}}^2Z_{\mu R}^{'}Z_{R}^{'\mu}.
 \label{eq:bosonmass1}
\end{eqnarray}

From Eq. (\ref{eq:bosonmass}), by following the standard model of electroweak interaction symmetry breaking procedure to the left-right symmetry model and put the value of $b=d$, we can see that the resulted gauge bosons masses for both left and right bosons are equal.  To make the left-right symmetry model relevant to the electroweak phenomena, many authors \cite{Mohapatra75, Senjanovic79, Mohapatra81} have taken the value of $b<<d$.  The difference values of the $b$ and $d$ can be associated with the parity violation.  

The gauge bosons masses problem in our model, especially for charged bosons masses arise from bidoublet Higgs vacuum expectation values, could be resolved by defining:
\begin{eqnarray}
\bordermatrix{&\cr
&m_{W_{L}}^{*} \cr
&m_{W_{R}}^{*} \cr}=\bordermatrix{& &\cr
&\cos\theta &-\sin\theta \cr
&\sin\theta &\cos\theta\cr}\bordermatrix{&\cr
&m_{W_{L}}(\phi) \cr
&m_{W_{R}}(\phi) \cr}
 \label{eq:mixing}
\end{eqnarray}
where the mass mixing angle $\theta$ could be associated with parity violation. From Eq. (\ref{eq:mixing}) we can see that the maximal parity violation occur for $\theta=45^{0}$ such that the total charged right bosons masses $m_{W_{R}}$ very large compared to the left bosons masses $m_{W_{L}}$.  For example, if we take values of the $W_{L}$ and $W_{R}$ bosons masses from doublet are $m_{w_{L}}=m_{w_{R}}=82$ GeV, then the reasonable values for gauge bosons masses from induced bidoublet are $m_{w_{L}}(\phi)=m_{w_{R}}=6724$ GeV or $m_{w_{L}}(\phi)=m_{w_{R}}(\phi)\approx 7$ TeV.  If we take $\theta=45^{0}$, then $m_{w_{L}}^{*}=0$ and $m_{w_{R}}^{*}=9509.17$ GeV.  Thus, the effective right charge-current contribution (V+A interaction) to the electroweak interaction is only around 0.0073 percent.  In this scheme, the domination of the V-A interaction over V+A interaction for charged current at low energy could be understood as an implication of the very massiveness of the $m_{W_{R}}$ with total mass $m_{w_{R}}=9591.17$ GeV $\approx 10$ TeV compared to $m_{W_{L}}= 82$ GeV.  The gauge bosons masses $m_{W_{L}}$ in the maximal mass mixing only come from the doublet Higgs field $X_{L}$ (as formulated in the GWS model) and supported by experimental fact that the boson masses come from doublet Higgs field.

\section{The Leptons masses}
As long as we know, for leptons masses there is no experimental fact to force us for choosing a one kind of multiplet Higgs in the Yukawa term except dictated by the requirement of the Lagrangian density must be gauge invariance.  But, in the boson mass sector, the representation of the Higgs fields to be $SU(2)$ doublet theoretically supported by experimental fact.  Thus, we could use the bidoublet Higgs field for generating the leptons masses.

Following GWS model, the leptons masses in our model also arise from the symmetry breaking.  The lepton mass term is the Yukawa term in Eq. (\ref{eq:completelagrangian}), that is:
\begin{eqnarray}
L_{l}=G\bar{\psi}_{L}\left\langle\phi\right\rangle\psi_{R}.
\end{eqnarray}
By inserting the vacuum expectation value of bidoublet Higgs, then we obtain the masses: $m_{\nu}=Gp$ and $m_{e}=Gs$ for neutrino and electron respectively.  For $p=s$, the neutrino and electron masses are equal.  Because the neutrino and electron masses arise from bidoublet Higss vacuum expectation values after symmetry breaking, then $m_{\nu}=Gp$ is a  mixing of the $m_{\nu L}(\phi)$ and $m_{\nu R}(\phi)$, and $m_{e}=Gs$ is a mixing of the $m_{eL}(\phi)$ and $m_{eR}(\phi)$.  Thus, in our minimal left-right symmetry model based on $SU(2)_{L}\otimes SU(2)_{R}\otimes U(1)$ gauge group the left-handed leptons masses arise from the bidoublet Higgs vacuum expectation values could be written as follow:
\begin{eqnarray}
\bordermatrix{&\cr
&m_{\nu L} \cr
&m_{\nu R} \cr}=\bordermatrix{& &\cr
&\cos\theta &-\sin\theta \cr
&\sin\theta &\cos\theta\cr}\bordermatrix{&\cr
&m_{\nu L}(\phi) \cr
&m_{\nu R}(\phi) \cr}
 \label{eq:mixingneutrino}
\end{eqnarray}
for neutrino sector, and
\begin{eqnarray}
\bordermatrix{&\cr
&m_{eL} \cr
&m_{eR} \cr}=\bordermatrix{& &\cr
&\cos\theta &-\sin\theta \cr
&\sin\theta &\cos\theta\cr}\bordermatrix{&\cr
&m_{eL}(\phi) \cr
&m_{eR}(\phi) \cr}
 \label{eq:mixingelectron}
\end{eqnarray}
for electron sector.

From Eqs. (\ref{eq:mixingneutrino}) and (\ref{eq:mixingelectron}) we can see that a tiny neutrino mass $m_{\nu L}$ could be obtained for mixing angle $\theta\approx45^{0}$, and the electron mass: $m_{eL}=m_{eR}$ for mixing angle $\theta = 0^{0}$.
\section{Conclusion}
By using the minimal content of the particles  (two primary doublets Higgs, one induced bidoublet Higgs (secondary Higgs), and two lepton fields) involve in the left-right symmetry model based on $SU(2)_{L}\otimes SU(2)_{R}\otimes U(1)$ gauge group, the domination of the V-A interaction over V+A interaction at low energy due to the very massiveness of the $m_{w_{R}}$.  The neutrino and the electron masses arise naturally, but for obtaining the leptons masses (neutrino and electron) phenomenologically, the symmetry breaking must be followed by mass mixing scheme.   The mixing angle $\theta = 45^{0}$ for boson sector, $\theta \approx 45^{0}$ for neutrino sector, and $\theta = 0^{0}$ for electron sector.  The parity violation in our model could be associated with the mass mixing in the bosons and leptons masses arise from the induced bidoublet Higgs via symmetry breaking. 
\section*{Acknowledgments}
The first author would like to thank to the Graduate School of Gadjah Mada University Yogyakarta where he is currently a graduate doctoral student, the Dikti Depdiknas for a BPPS Scholarship Program, and the Sanata Dharma University Yogyakarta for granting the study leave and opportunity.

\end{document}